# QUANTUM MAGNETISM AND SUPERCONDUCTIVITY

by William J.L. Buyers and Zahra Yamani

The spin of the neutron allows neutron scattering to reveal the magnetic structure and dynamics of materials over nanometre length scales and picosecond timescales. Neutron scattering is particularly in demand in order to understand high-temperature superconductors, which lie close to magnetically ordered phases, and highly correlated metals with giant effective fermion masses, which lie close to magnetic order or pass through a mysterious phase of hidden order before becoming superconducting. Neutron scattering also is the probe of choice for revealing new phases of matter and new particles, as seen in the surprising behaviour of quantum spin chains and ladders where mass gaps and excited triplons replace conventional spin waves. Examples are given of quantum phenomena where neutron scattering has played a defining role that challenges current understanding of condensed matter.

**In this article, examples are given of quantum phenomena where neutron scattering has played a defining role that challenges current understanding of condensed matter.**

## INTRODUCTION

Magnetism is at the heart of fundamental processes. The way in which black holes suck in matter from neighbouring stars is a fundamentally magnetic process and not just caused by gravity[1]. The magnetic moment of the neutron allows scientists to study magnetism in materials at the nanoscale and below. The pattern of neutron scattering and its velocity distribution reveals the structure and dynamics of the atomic magnetic moments or spins. Spins in condensed matter belong to the overall electron system and arise from unpaired electrons in one of the outer orbital shells of the atom. The ground state of the spin depends on whether it is surrounded by and exchange coupled to other well-defined spins as in insulators, or whether the spins are embedded in a liquid of conduction electrons that may screen their moment and damp out their excitations.

In insulators an integral charge state is determined by chemical valency and the environment allows the several unpaired electrons to form localized states of definite orbital and spin angular momentum linked by spin-orbit coupling. Hund's rule is king. At each site we have an independent atom the symmetry of whose orbit is lowered by the electrostatic field of its neighbours. The Pauli exclusion principle leads to an effective coupling, J, between neighbouring spins that may be ferromagnetic (parallel) or antiferromagnetic (antiparallel). The latter is more prevalent in nature because the magnetic atoms in insulators establish superexchange bonds through shared non-magnetic neighbours such as $O^{2-}$ in MnO or $F^-$ as in $KMnF_3$. The atom behaves magnetically as if it has a fixed magnetic moment that precesses around the sum of the static and dynamically varying field of its neighbours. This is the site-based picture of localized spins.

In metals the situation is entirely different. The conduction arises from band-based electrons in which it is the electron momentum that is well defined at the Fermi surface rather than the electron position. Although embedded in a liquid of high-velocity conduction electrons, local spins may still behave independently provided their energy scale, given by the exchange coupling, J, is much less than the eV bandwidth of the fast conduction electrons. The conduction electron spin responds adiabatically to the motion of the slow local spins. This is the picture for the rare earth metals, except for a few mixed-valent examples. The decoupling works because the small-radius 4f magnetic shell lies inside the 5s shell and so is shielded from the destructive influence of its neighbours. Even in this weakly coupled system the spin excitations of the f electrons are not eigenstates - the indirect coupling through the conduction electron sea shortens their lifetime. They acquire a relaxation rate, seen as a spectral line width, proportional to the imaginary part of the conduction electron (Lindhard) spin susceptibility $\chi''(\mathbf{q},\omega)$ because coupling, I, of the local f-spin to the conduction electron spin causes an indirect (RKKY) exchange between f-moments of the form $J(\mathbf{q},\omega)=I^2\chi(\mathbf{q},\omega)$. The same indirect coupling, now through the charge susceptibility, gives phonons in metals a spectral broadening, and this is only removed for energies below the pairing gap in its conduction electron charge response when the metal becomes superconducting below $T_c$ [2].

Nonetheless, when the coupling to conduction electrons is strong by exchange or by hybridization, the spins behave as if they are free at high temperature but are progressively screened on cooling by coherent reorganization of the conduction electrons. The effect is described as Kondo screening when the spin of the atomic core and of the conduction electron can reorganize without substantial change of charge state and is described as mixed valency when hybridization changes the occupancy and effective charge.

William J.L. Buyers <william.buyers@nrc-cnrc.gc.ca> and Zahra Yamani, Canadian Neutron Beam Centre, National Research Council, Chalk River, Ontario, Canada K0J 1J0

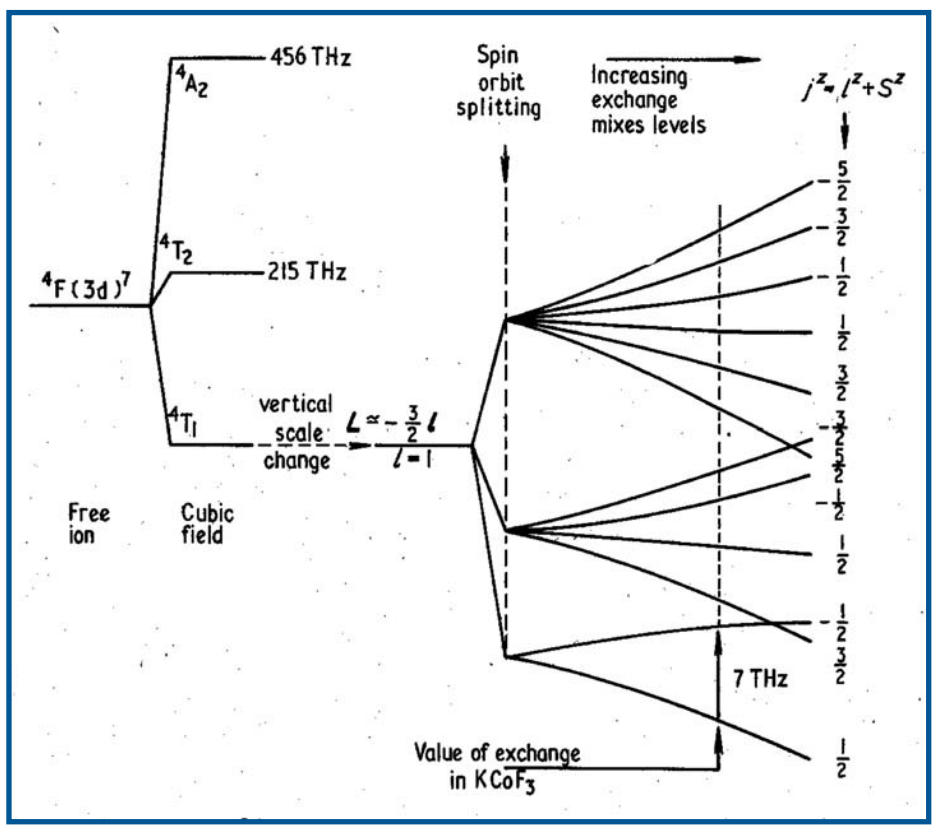

Fig. 1  Energy levels of Co$^{++}$ in the antiferromagnet KCoF$_3$. The spin waves are transitions from the ground to all excited states and form a lowest band up to the illustrated single-ion spin-flip frequency of 7 THz and down to a gap frequency set by the exchange mixing of higher spin-orbit states [3].

Examples of exotic or unconventional phenomena discovered with neutron scattering include magnetic solitons [4,5], and the quantum gap (Haldane gap) between the ground state and a triplet of massive spin particles that appears for integer but not half-integer spin chains. One-dimensional chains of spins are created by chemically separating chains of magnetic ions by ligands of non-magnetic ions. Clever solid-state chemists are responsible for creating the wide variety of 1D, 2D and 3D magnetic systems where singlet ground state, spin liquid and quantum phenomena may be investigated.

Solitons in an Ising-like antiferromagnetic chain, where the spins point up or down, must be excited in pairs by the neutron flipping a spin ($S_z \rightarrow -S_z$), thereby creating *two* domain walls or solitons costing energy 2J because there are two wrong bonds with ferro- instead of antiferro-orientation. This is because neutron scattering only connects states linked by the spin operator – it is a spin-one probe. The Ising spin exchange $JS_i^z S_{i+1}^z$ in the presence of weak transverse coupling allows each soliton to hop two sites at a time away from the initially localized soliton pair (Fig. 2). These solitons can be visualized as a place where we have twisted the rest of the chain through 180° to make a π soliton. Because the initial excitation carried spin one, we find that each soliton is a spin one-half particle. A single soliton may be thermally excited with an activation energy J, half the spin wave energy. This simple example from the Ising chain has given rise to the concept of spinons, the basic particle in the S=½ isotropic (or Heisenberg) chain, later used for high-temperature superconductivity. Because they are created in pairs, conservation of momentum ensures that there is a continuum of spinon excitations instead of sharp spin waves, a continuum that extends down to a lower limit set by the Bethe ansatz.

In pure metallic systems spins normally condense into a state whose symmetry is lowered as a result of formation of magnetic order, a spin-density wave, a charge density wave, or a superconducting paired state, while some systems remain paramagnetic to the lowest temperatures.

A spin or orbital excitation appears as a collective excitation of the ordered state whose energy-momentum dispersion relation is a direct measure of the magnetic forces between any two atoms. Their energies give information on the local crystal field, the spin-orbit coupling and the interatomic exchange as shown in Fig. 1 for the insulator KCoF$_3$. Although it has become customary in orbitally ordered materials such as manganites to treat the orbitons separately from the spin waves, spin wave and orbital states are not distinct as they are coupled by spin-orbit interaction. They together form the collective magnetic dipole excitations of the electronic system and should be included in an extension of the standard model [3].

### EXOTIC PARTICLES

More than just measuring the strength of interactions, as may be done in well-understood systems where the magnetism appears in the form of well-defined spin waves, the neutron is uniquely suited to discovering new phenomena that are not contained within accepted textbook lore.

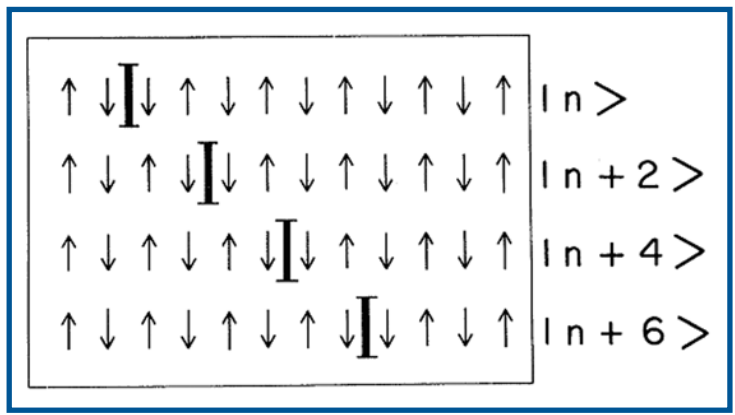

Fig. 2  Solitons hopping along a chain of S=½ spins [5]. Because the excitation is topological (half the spin chain is turned over at each thickly marked wall) as opposed to the sinusoidal spin-wave pattern, the soliton response looked at with a Fourier probe such as neutron scattering appears as a continuum.

When the spins interact with isotropic rather than the above Ising exchange, classical thinkers, and most scientists, expected that the spin spectrum would be gapless. Instead Haldane conjectured [6] that the chains of integral-spins would exhibit a mass gap but those of half-integral spins would not. The Haldane gap was not expected, at least not by those that wave their fingers to illustrate rotational invariance and at the same time consider a long wavelength spin wave to be a precession of a classical spin! Before Haldane, Kadanoff's postulate of Universality had wide acceptance because both experimental results on different materials and theoretical calculations with $S=\frac{1}{2}$ and $S=\infty$ (i.e., classical result) gave phase transitions with the same properties. Renormalization group theory offered a mechanism for universal properties to arise since under repeated renormalization transformations some parameters are attenuated and become irrelevant while others remain relevant. Haldane's conjecture was controversial mainly because it contradicted Universality. While it is sometimes valid to take a site-based view where the spin precesses in the field of its neighbours, this largely works only when there is a static field (a magnet with long-range order), in high dimensions, and in lattices without frustration (competing exchange fields). In one dimension there can be no long-range order and the spins can attempt to form bond order where pairs of spins form a singlet ground state. These singlet pairs may then interact and it is not obvious *a priori* whether this will give a lower overall ground state than the site-based approach. Anderson showed that the 2D triangular lattice preferred a ground state of resonating valence bonds over the Néel state, but the situation for one dimension (1D) was unknown until the work of Haldane [6].

The Haldane conjecture remained controversial, as well as being counter-intuitive, until the spin gap was discovered directly in neutron scattering experiments at Chalk River involving a University of Toronto student [7]. The isotropically coupled S=1 $Ni^{2+}$ chains in $CsNiCl_3$ were the test bed. The conventional (linear spin wave theory) view was that the lowest spin excitations were gapless Goldstone bosons but as shown in Fig. 3 the integral $Ni^{2+}$ spins exhibit a large gap, about 40% of J. This result was soon confirmed in Europe on an organic material [8] and in $CsNiCl_3$ polarized neutron scattering showed that the gap states were triplets [9].

In recent years the unusual temperature dependence of the gapped triplet states, which has given rise to the new name for a particle, the triplon, have been fully explored by Kenzelmann *et al.* [10]. The spin triplons increase their energy on heating, whereas spin-wave energies decline in ordered systems (Fig. 4). Within the non-linear sigma model, this is because, to conserve the total moment, the triplon energies must rise to counteract their increase in population through thermal excitation.

A useful picture for a singlet with a gap to excited states is the valence bond solid described in the review by Affleck [11], in which each of the two electrons of an S=1 atom form a singlet pair with one electron of a neighbouring atom, one to the left and one to the right. This global singlet state is the exact solution of a closely related Hamiltonian. For $S=\frac{1}{2}$ the sole electron can form only one singlet bond, all to the left or all to the right, but then there are two degenerate states, no singlet and no gap.

The discoveries of the quantum gap presaged the large current body of research on singlet-to-triplet excitations or triplons, as found in even-leg spin ladders and in systems formed from integer-spin triangle motifs. The solitons in magnetic chains led to the now pervasive modern concept of spinons.

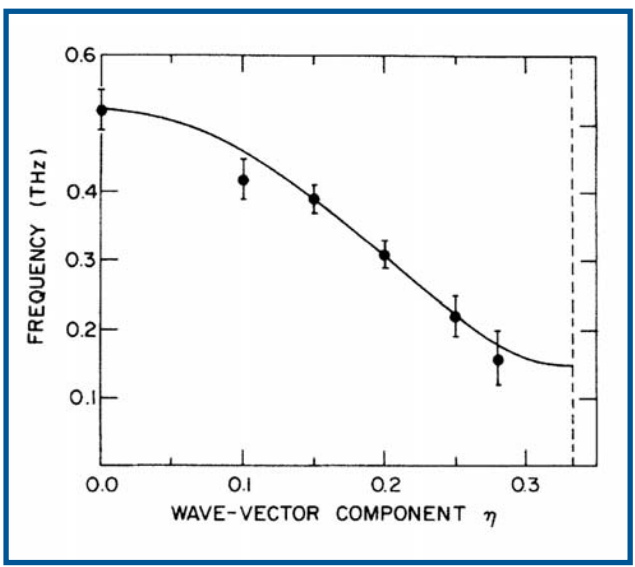

**Fig. 3** The Haldane spin gap discovered in the integer-spin chain system $CsNiCl_3$ in its 1D phase [7]. The $Ni^{2+}$ (S=1) chains lie along the hexagonal z direction [0 0 1]. If the excitations were conventional spin waves all frequencies along the 1D zone centre Q =($\eta,\eta$, 1) would lie at zero, since there is no long-range order, but the quantum disordered ground state leads to a mass gap of 0.32 THz to a triplet of spin excitations with only short-range spin correlations. The weak coupling perpendicular to the chains along ($\eta,\eta$, 0) leaves a residual in-plane dispersion.

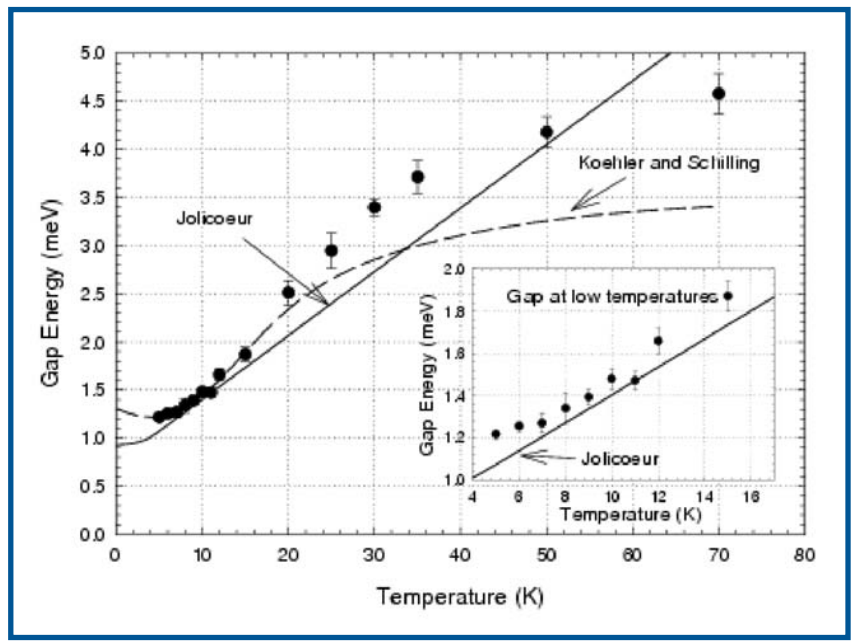

**Fig. 4** Haldane gap triplet energies rise with temperature in accordance with the self-consistent non-linear sigma model [10].

**HIDDEN ORDER**

An enigmatic problem in the field of strongly correlated heavy fermion systems is the nature of the hidden order that sets in below the large specific heat jump at $T_0 = 17$ K in $URu_2Si_2$. In addition, a superconducting phase occurs in $URu_2Si_2$ below 1.2 K. The heavy-fermion epithet stems from the fact that the Sommerfeld specific heat coefficient, $\gamma=C/T$, usually taken as a measure of the electronic density of states at the Fermi surface, is large above $T_0$, 160 mJ/mol-K$^2$. This is a hundred times larger than that of a simple metal like copper and suggests that the effective electron band mass is a hundred times the free electron mass. Clearly the proximity to a magnetic or exotic transition is causing the large mass, through spin or hybridization effects. Since evidence of heavy charge masses has been seen in de Haas-van-Alphen experiments, the strong spin response must be producing a slowing of the electron velocities, although the spins can themselves contribute to the giant specific heat. The charge and spin spectra must be renormalized downward to a few meV in energy to add to specific heat.

Although a second order transition occurs at 17 K with substantial associated entropy, the nature of the order has remained a mystery for over 20 years [12,13]. Landau theory tells us that the small antiferromagnetic moment of 0.03 $\mu_B$ that develops below 17 K cannot possibly explain the large specific heat jump (entropy) associated with a second order local spin transition. It would require ordering of a moment ~100 times larger! Antiferromagnetism therefore cannot be the hidden order parameter. The system seems rather to have condensed into a new phase of matter for which the order parameter and associated symmetries differ from conventional expectations. The properties are typical of ordering due to broken symmetry but, since its origin has not yielded to practically every known experimental technique, we refer to it as 'hidden' order. Strong hybridization is expected between the conduction and the 5f electrons and prevents application of either a purely localized or itinerant electron model. Local probes and pressure experiments suggest that the weak moment may be a parasitical phenomenon that forms in a very small volume fraction. The small moment may be simply a quixotic distraction from the real bulk order parameter that causes the large loss of electronic density.

What is clearly a bulk property of the hidden order phase is the unusual spectrum of magnetic excitations (Fig. 5). Neutron scattering has shown [13] that they form well-defined propagating collective modes over most but not all of the Brillouin zone. Moreover they carry a large spin matrix element of 1.2 $\mu_B$ and are thus a property of the bulk or dominant phase. What is unusual is that the spin motion is entirely longitudinal along the tetragonal c direction. Contrast this with a spin wave of a magnetically ordered system where the motion is transverse to the moment. Also unusual is that while the well-defined excitations suggest a localization of the 5f moment, along the tetragonal [0 0 1] direction the lifetime shortens and damps out the response and so suggests decay into an itinerant-electron continuum. Itinerant spins are also suggested by long-range (RKKY) exchange that produces the several extrema in the dispersion relation. Over the last two decades many searches have been carried out for the hidden order and the evidence is either absent or contrary to models involving charge-density wave formation, quadrupolar ordering, multispin correlators [14], crystal fields [15], or orbital currents [16].

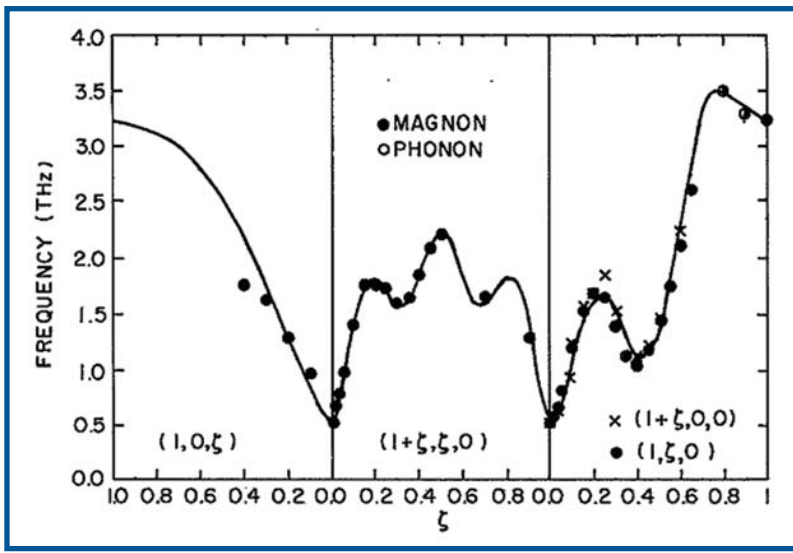

**Fig. 5** The frequency of gapped spin excitations versus wave vector in $URu_2Si_2$ at 4 K well within the hidden order phase [13]. Long-range exchange through the electron liquid causes several minima with the minimum gap at Q = (1, 0, 0). For directions within the tetragonal basal plane the excitations are long-lived, but those propagating in the c direction along (1, 0, $\zeta$) are damped out at large wave vector.

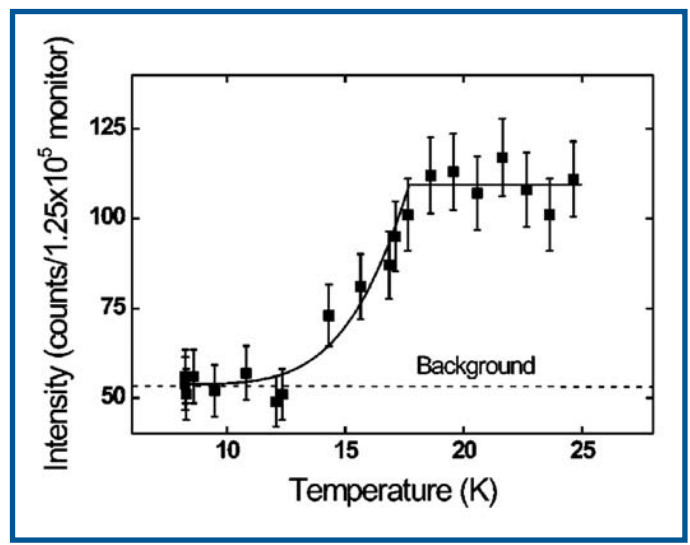

**Fig. 6** The temperature dependence of the incommensurate fluctuations at (1.4, 0, 0) and $E$=0.25 THz (~1 meV) energy transfer [17]. The fit gives an activation temperature of 110±10 K, the coherence temperature for the charge transport not the spin excitation energy.

A significant new result came from a search for an exotic form of magnetism predicted to arise below $T_0$ from orbital currents where the weak moment would result from electron currents flowing around the atoms in a unit cell [16]. The orbital moment theoretically predicted was not observed. Wiebe et al [17] made a more important observation, however. Above the 17 K transition the spectral weight moves to the incommensurate wave vector (1.4, 0, 0) of the second minimum (Fig. 5) and the spectrum becomes gapless. The onset of the collapse of the gap was measured by probing the fluctuations at 0.25 THz, well below the gap frequency of 1.2 THz. The important result is that the incommensurate scattering is activated with a temperature $T^* = 110$ K, the coherence temperature (see Fig. 6). Thus in the precursor phase to hidden order there are gapless incommensurate spin fluctuations over a finite region of the Brillouin zone. These can give rise to a term in the specific heat linear in T that can be misconstrued as electronic specific heat. The specific heat will jump at $T_0$ and decrease below as the spin gap is formed. The large linear-in-T specific heat then may be thought of as coming from the spin fluctuations rather than from a "heavy-electron" charge band. Theorists often like to work with the "one band does all" approach with a Hubbard model that tries to reproduce both the charge and the spin response. Whereas most focus on the fermions determining the charge transport properties, the new results require more attention to the bosons of the spin response. The hidden order phase is robust and persists to a field of 35 T [18]. These exotic results have led to exotic theories, most recently to the suggestion that a Pomeranchuk instability of the electron liquid has grossly changed the Fermi surface [19].

## SUPERCONDUCTORS

Neutron scattering is particularly well suited to explore the intimate relation between magnetism and high-temperature superconductivity. The superconductors consist of square-lattice $CuO_2$ planes of $S=½$ copper spins, into which holes have been created in the plane by the chemical removal of electrons from oxygen ligands, a process known as doping. Neutron scattering can show how the spin spectrum evolves as the superconducting transition temperature increases and then decreases as the electronic doping is increased beyond a critical value, $p_c \sim 5\%$, into the phase known as the superconducting dome. The $S=½$ holes sit equally on the four oxygen neighbours and, from a distance, screen the copper moment to form a spin singlet. The resonating valence bond (RVB) ground state has been adduced to account for the precursor state that connects a Mott insulator ($LaCuO_4$ or $YBa_2Cu_3O_6$) to the hole-doped state where high-temperature superconductivity takes place [20]. The RVB state consists of sets of singlet pairs between copper spins at all distances with a symmetry similar to that of a superconducting pair. In conventional (phonon or S-wave) superconductors the pairing gap occurs for all directions of Fermi momentum, $k_F$. In contrast the spins of the RVB pair lie on different atoms and the gap has d-wave symmetry with nodes along the directions $k_x = \pm k_y$.

Although there is a large amount of neutron beam research on $La_{2-x}Sr_xCuO_4$ and $YBa_2Cu_3O_{6+x}$, most is for relatively highly doped materials. In recent years attention has shifted in three continents to underdoped materials where superconductivity is weaker but magnetic fluctuations are stronger [21,22,23]. With the advent of high quality crystals from University of British Columbia it has been possible to study highly-ordered ortho-II crystals that display greater electronic order and thus a larger $T_c$ for the same oxygen doping. With these crystals the hour-glass spectrum of incommensurate spin modulations at low energy, a resonance localized in Q and in ω, and a cone of damped high-energy spin waves has been well-established in recent work at Chalk River and at ISIS in the UK [24,25].

Here we focus on systems that lie much closer to the critical onset of superconductivity where the destruction of spin order and spin wave propagation seems the most crucial requirement for the onset of this anisotropic superconducting charge pairing. A recent study [26] has shown that high-temperature superconductors close to the edge of the superconducting dome behave quite differently from both their more highly doped counterparts and from their antiferromagnetic parent compounds. Although no Bragg peak, and so no long range order, is observed for lightly doped superconductors, subcritical 3D antiferromagnetic correlations are formed. This is evident from fact that the spin scattering is centred at integer values of L for zones (½,½,L) (Fig. 7 ). Thus the doped holes prevent the formation of the long range ordering but there is a memory of the phase that would be formed by further reduction of the hole content.

Compared to the higher doped materials with a high-energy resonance (33 meV for YBCO6.5) at a commensurate position and no elastic central mode [25], the energy spectrum of lightly doped superconductors consists (Fig. 8) of a central mode coupled to a broad inelastic peak with a relaxation of 2.5 meV. Both are centred on the commensurate antiferromagnetic position but are broad in momentum. Correlation lengths associated with both modes are short ranged (longer in the basal plane than along the c-axis). The intensity of the central mode increases on cooling from 80 K and saturates at a low temperature of order of 10 K with no suppression at $T_c$ (Fig. 9).

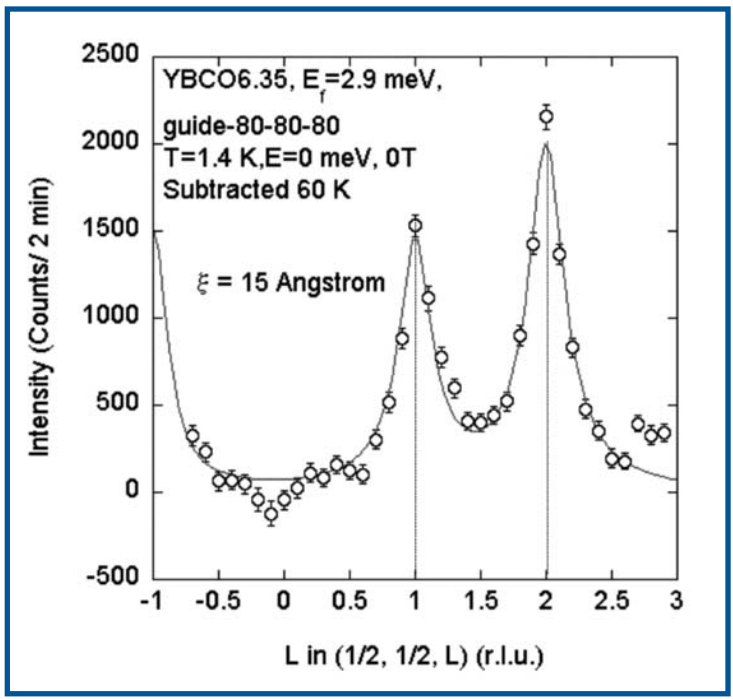

Fig. 7    In YBCO6.35 with $T_c$=18 K, the antiferromagnetic correlations coupling the planes extend over only 15 Å along the c-axis [27].

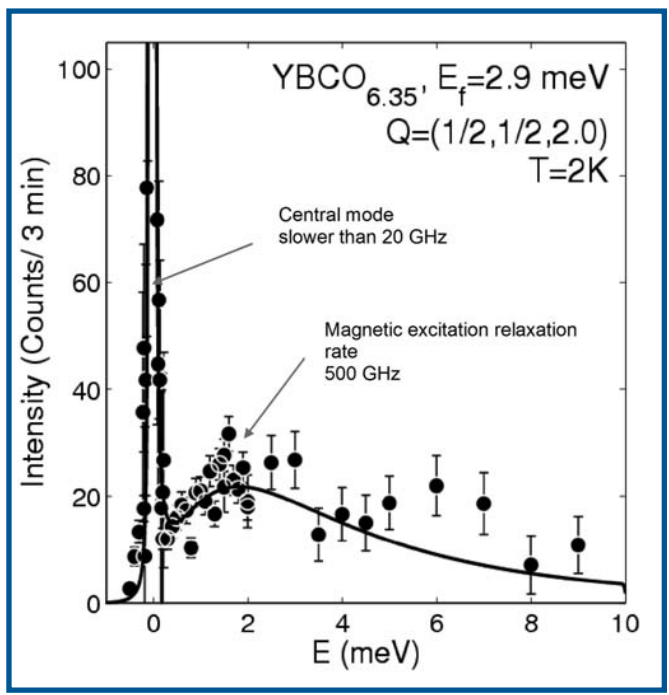

Fig. 8  Two magnetic energy scales near the onset of superconductivity in $YBa_2Cu_3O_{6.35}$, a narrow central mode with FWHH<0.08 meV, and faster relaxational excitations peaked at ~2 meV. The line is from a model where the soft relaxational magnetic mode of the superconducting phase is coupled to an elastic (central) mode and drives up its intensity to divergence as a quantum phase transition to the ordered magnetic phase is approached [26]. The nearly-elastic mode arises from the slow tumbling of about a hundred copper spins that are nearly ordered.

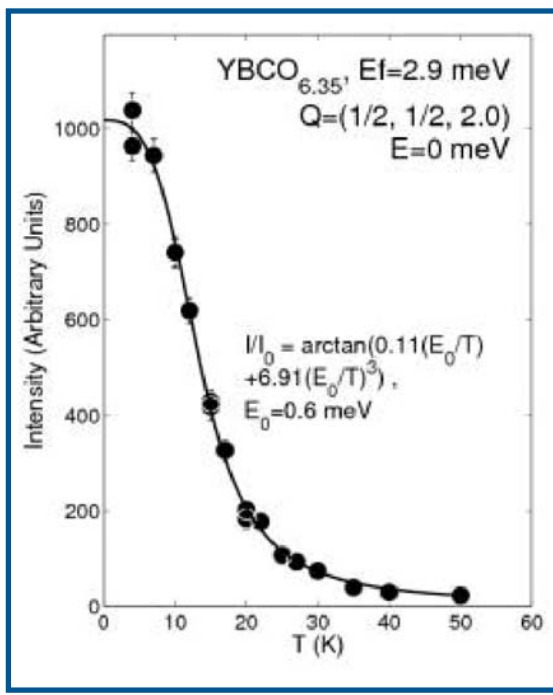

Fig. 9  The central peak grows on cooling with no change at $T_c$ = 18 K as if the spins ignore superconducting transition.

The general behaviour as holes are added is that the strong antiferromagnetism of the parent insulators is rapidly broken up, carriers form to conduct electricity and heat, and the spin excitations evolve into strongly damped paramagnons [25]. Long-range antiferromagnetism has been destroyed and a superconducting phase is entered with only ~5% of hole doping. This is much less than the percolation limit of ~50% localized vacancies for a dilute 2D lattice, and clearly shows that holes produce a large spatial extent of weakened AF coupling. Possibly local ferromagnetic correlations ensue (Fig. 10).

Perhaps the most surprising property, observed with polarized neutron scattering [26], is that the spin orientation is isotropic, unlike the XY order of the insulator. We can infer that the superconductor is in a spin 'hedgehog' phase. Such preservation of spin rotational invariance is a very different topology than the collinear spins of the antiferromagnetic insulator. It suggests a frozen glass state that inhibits the transition to magnetic long-range order and provides the random spin environment that allows superconductivity.

One of the most remarkable features of the cuprate superconductors is the characteristic spin excitation energy known as the "resonance". It tracks $T_c$ as the doping is varied (Fig. 11). The spin spectrum exhibits a peak whose energy scales as $E_{res} \sim 6k_BT_c$. Inclusion of the results [26] for p=0.06 ($YBa_2Cu_3O_{6.35}$) shows that the inelastic spin energy, albeit reduced by an order of magnitude in energy from that of optimally doped YBCO and heavily damped (Fig. 8), is a critical spectral feature of superconductivity. Fig. 11 shows it is the soft mode of the superconducting dome.

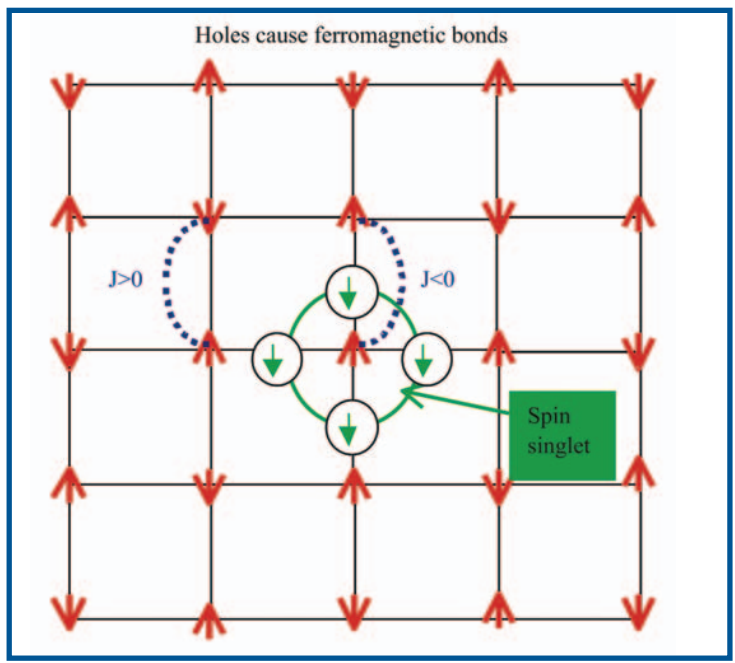

Fig. 10  A doped hole on the oxygen neighbours puts the $CuO_4$ into a singlet state and may cause ferromagnetic bonding. Even a low doping of ~5% destroys the antiferromagnetic order because every hole affects many sites.

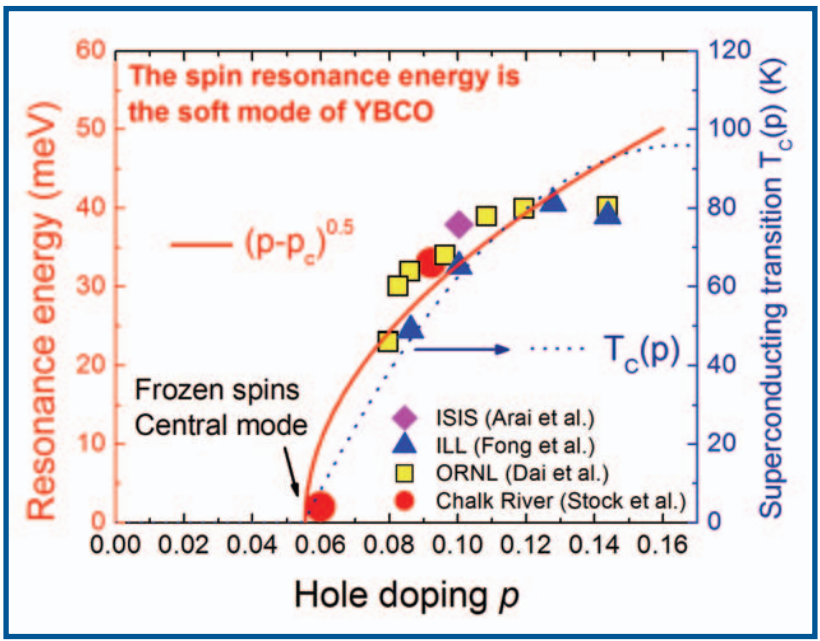

**Fig. 11** The characteristic energy of the inelastic spin response tracks the superconducting transition temperature $T_c(p)$ as the doping p in the $CuO_2$ planes is increased.

In cuprate superconductors the reason for the remarkable tracking of the superconducting transition temperature with the resonance energy (Fig. 11) has not been explained satisfactorily. While a number of theories based on a Fermi liquid coupled to a spin susceptibility have attempted to explain individual experiments at large, near-optimum doping where the electrons form a Fermi liquid, these theories are unlikely to work in the region near the lower edge of the superconducting dome that we have studied. There the electronic hole density is small, there is considerable doubt as to whether a sharp Fermi surface exists, and the resistivity is insulator-like, falling with increasing temperature, thus mirroring the decrease in $\chi''(\mathbf{q},\omega)$ with frequency. Moreover the spin fluctuations are so strong (recall from Fig. 9 they ignore $T_c$) that a description based more on states, RVB or otherwise, that pre-exist at a less-than-critical doping would seem a better starting point.

In this regard we suggested that the resonance can be regarded [24] as an image of the two-particle pairing states, states that are allowed in the particle-hole spectrum detected by neutrons only by dint of the superconducting order. Because the pairing gap is d-wave of the form $\cos(k_x)-\cos(k_y)$, the spectrum of spin states coupled incoherently to all electron momenta should exhibit an anisotropic rise to a peak at the maximum d-wave gap followed by a sudden fall. This asymmetric resonance spectrum is very close to what was observed in an oxygen-ordered crystal of YBCO6.5 (Fig. 12 based on [24]) and may be a fingerprint of superconducting pairing. Moreover, in the normal phase almost half the resonance weight has already formed on cooling to just above the superconducting transition temperature. We believe that this fingerprint shows that incoherent superconducting pairs are present in the normal phase. By contrast conventional phonon-mediated superconductors show an extremely nar-row temperature range for critical fluctuations. When we reduce the doping to the edge of the superconducting phase we have seen that the spin fluctuations are strong and in this interpretation are dominated by incoherent pairs, so much so that they show little change in their growth rate on cooling through $T_c$. Needless to say this concept is highly controversial, for it would suggest that the lightly-doped but non-superconducting antiferromagnet would carry some of the same local pairing symmetry as the superconductor.

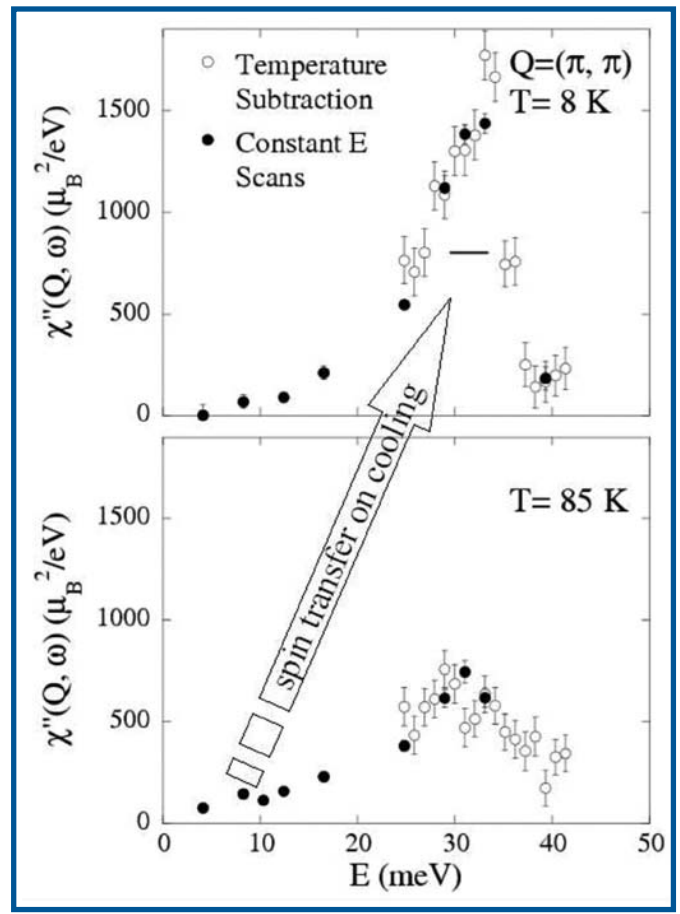

**Fig. 12** The spin resonance peaked at 33 meV in YBCO6.5 in its superconducting phase (8 K) and in its normal phase (85 K) above its superconducting transition temperature of 59 K. The two-dimensional wave vector $(\pi,\pi)$ selects spin fluctuations that have opposite sign (are of antiferromagnetic symmetry) between neighbouring Cu atoms in the square lattice. For this AF phasing there is no gap in an ordered antiferromagnet. In the superconductor with its d-wave gap for pairing charge carriers, the spin response is shifted upward. The presence of a similar but weakened spectrum above $T_c$ indicates that local incoherent pairs have already formed in the normal phase [24] within vortex-antivortex fluctuations.

## CONCLUDING REMARKS

The power of neutron scattering is that it provides direct access to the energy, momentum and spin of the fundamental particles in condensed matter systems. Other spectroscopic techniques are generally less direct, such as the local probes of muon spin resonance and NMR. The positive muon traps and interacts strongly on the large eV scale with its immediate electronic environment drawing a screening electron around it; the field it measures may in some systems be different on the meV scale of spin fluctuations than the unperturbed field of the system. Other probes give an average of the charge but not spin spectra, or are averages over many particles such as thermal and electrical conductivity and specific heat. Neutron scattering has allowed new phases of matter to be discovered as we have seen for quantum gapped systems, for a highly-correlated heavy-fermion system, and for the quantum antiferromagnet doped to form a superconductor.

## ACKNOWLEDGEMENTS

WJLB benefited as a member of the Canadian Institute for Advanced Research and both authors recognize technical and scientific support from CNBC, Chalk River, and NIST, Gaithersberg, MD. We are grateful to C. Stock, and to many colleagues, for their insight and help over several years.